\documentclass[superscriptaddress,showpacs,amsmath,amssymb,prb,twocolumn]{revtex4}
\usepackage{graphicx}
\usepackage{dcolumn}
\usepackage{bm}
\usepackage{color}
\usepackage{amsmath}
\usepackage{amssymb}
\usepackage{epstopdf}
\usepackage{svn-multi}
\svnid{$Id: ePDF.tex 7793 2014-01-08 22:17:43Z juhas $}
\graphicspath{{./figures/}}

\def\sjba#1{}
\def\nba#1{}               

\begin{document}

\title{Calibration and data collection protocols for reliable
 lattice parameter values in electron pair distribution function {(ePDF)} studies}
\author{A. M. Milinda Abeykoon}
\affiliation{Condensed Matter Physics and Materials Science Department, Brookhaven National Laboratory, Upton, New York 11973, USA\\}

\author{Hefei Hu}
\affiliation{Condensed Matter Physics and Materials Science Department, Brookhaven National Laboratory, Upton, New York 11973, USA\\}

\author{Lijun Wu}
\affiliation{Condensed Matter Physics and Materials Science Department, Brookhaven National Laboratory, Upton, New York 11973, USA\\}

\author{Yimei Zhu}
\affiliation{Condensed Matter Physics and Materials Science Department, Brookhaven National Laboratory, Upton, New York 11973, USA\\}

\author{Simon J. L. Billinge}
\affiliation{Condensed Matter Physics and Materials Science Department, Brookhaven National Laboratory, Upton, New York 11973, USA\\}
\affiliation{Department of Applied Physics and Applied Mathematics, Columbia University, New York, NY 10027, USA}
\email{billinge@bnl.gov}

\begin{abstract}
We explore and describe different protocols for calibrating electron pair distribution function (ePDF) measurements for quantitative studies on nano-materials.  We find the most accurate approach to determine the camera-length is to use a standard calibration sample of Au nanoparticles from National Institute of Standards and Technology. Different protocols for data collection are also explored, as are possible operational errors, to find the best approaches for accurate data collection for quantitative ePDF studies.
\end{abstract}

\date{\today}

\pacs{, }
\preprint{Draft manuscript for submission}

\maketitle

\section{Introduction}
In an earlier paper~\cite{abeyk;zk12} it was demonstrated that atomic pair distribution functions (PDFs) obtained from electron diffraction data from nanoparticles, collected in a standard transmission electron microscope (TEM), could be modeled to obtain quantitatively reliable structural information about the 3D structure, opening the possibility that the electron-PDF or ePDF method may become a convenient laboratory based nanoparticle structural characterization approach.  In this way the powerful atomic PDF methods~\cite{egami;b;utbp12,billi;cc04,billi;jssc08,young;jmc11} may be opened to a much broader audience since, currently, most data for PDF studies are collected at large user facilities such as X-ray synchrotrons and neutron sources where access is an issue.  With ePDF it is possible, rather easily, to get PDFs using equipment that is already available in many chemistry and materials science laboratories, allowing PDF to become part of the standard suite of sample characterization tools.  Furthermore, quantitatively reliable PDFs may be obtained from very small amounts of material, micro or nanograms of nanoparticles or thin films~\cite{cocka;armr07} and still be analyzed using modern PDF modeling tools~\cite{farro;jpcm07,neder;b;dsadss08,cerve;jac10}.  The earlier paper~\cite{abeyk;zk12} demonstrated proof of principle; however, a number of experimental issues need to be addressed before this can become a standard technique and we explore a number of these issues here.

Similar to rapid data acquisition X-ray PDF (xPDF) experiments using
area detectors~\cite{chupa;jac03},  in ePDF measurements  the effective sample to detector distance, or camera length, must be determined to convert the position of pixels on the detector from geometric units to scattering angle, $2\theta$, or the magnitude of the momentum transfer, $Q$ or $s$, where $Q=s=4\pi\sin\theta/\lambda$ and $\lambda$ is the electron wavelength (Here $Q$ and $s$ are commonly used in the PDF and electron diffraction literature, respectively, for the same quantity~\cite{abeyk;zk12}).
In xPDF studies the wavelength of the
incident radiation is determined first, and the diffraction pattern of a calibration sample of known lattice parameter is measured.
The Debye-Scherrer rings from this pattern are then fit
with the known value of the lattice parameter fixed but allowing the camera length to vary. Once the camera length and wavelength are known, the calibration is replaced with the sample of interest (SOI) which is measured without changing any of the experimental setup. The camera length and wavelength are then transferrable to the  data from the SOI. In structure refinements on these data the known calibration constants are then fixed but the lattice parameter of the SOI varied in the refinement~\cite{egami;b;utbp12}.

In xPDF measurements, this same calibration
dataset is also used to determine any non-orthogonality between the incident beam and the
detector~\cite{hamme;esrf04}, and in determining instrumental resolution parameters such as $Q_{damp}$ in PDFgui~\cite{farro;jpcm07}.

These procedures are well established for xPDF measurements.  However, a number of issues need to be addressed in the context of ePDF.
First, a well characterized calibration sample for the ePDF case
is problematic as the method requires the study a nano-material~\cite{abeyk;zk12}, but the lattice
parameters of nanoparticles are often modified from the corresponding bulk material~\cite{gilbe;s04,masad;prb07}.  Here we explore different options for ePDF calibration samples.
The calibration process is further complicated in ePDF because unlike xPDF the camera length of the instrument depends on easily varied parameters such as  accelerating voltage and the magnetic lens settings, as well as the precise location of the sample in the instrument. We also explore here the error in the calibration that is brought about by incorrect procedures in this process, and suggest an optimal protocol for calibration and data collection in an ePDF measurement.

\section{Experimental methods}

All electron diffraction data presented in this paper were collected
using a JEOL ARM 200CF transmission electron microscope equipped with two sets of objective
apertures, the upper objective aperture (UOA) located in the diffraction plane and the lower
objective aperture (LOA) located below the diffraction plane. All electron diffraction patterns were recorded on a Gatan Orius CCD that eliminates charge overflow to the neighboring CCD pixels due to center beam saturation, and no beam-stop was used since charging of the beam-stop can introduce aberrations in the diffraction pattern~\cite{abeyk;zk12}. An accelerating voltage of 200~keV
was used with a camera length of $\sim80$~mm yielding a usable range of $Q$ up to
$Q_{max}=23$~\AA$^{-1}$.

Two possible standard samples were considered: a standard Au nanoparticle sample and a
commercially available evaporated aluminum film.
The standard Au nanoparticle specimen was obtained from the National Institute of Science and Technology (NIST)
in Gaithersburg, MD, US and has the designation SRM8011.  It consists of Au nanoparticles that are
nominally either 10~nm or 30~nm in size (depending on which standard is selected), suspended in solvent.
The evaporated aluminum film deposited on G400, 400 square mesh copper
Gilder grids was obtained from Ted Pella,~Inc. and is a standard TEM camera length
calibration sample.

All xPDF studies presented in this paper were carried out at the
beamline X17A at National Synchrotron Light Source (NSLS) at the
Brookhaven National laboratory. A commercial Ni powder sample from Alfa Aesar (catalog number: 10674) was
used to calibrate both sample-to-detector distance and the instrument
resolution parameter, $Q_{damp}$~\cite{egami;b;utbp12} of the xPDF setup, which was
then used to determine the lattice parameters of the Au nanoparticle
standard sample.

Integration of 2D electron diffraction was done using
the software package Fit2D~\cite{hamme;hpr96}, which has built-in features for
doing the calibration.  The most convenient uses a small set of internal standard samples
whose structural information is known by the program.  It is also possible to
use a user-defined calibration sample as we had to do here, because the $d$-spacings
of our calibration nano-samples are not necessarily the same as those of the same
material in the bulk. To create a user-supplied calibration file the first few $d$-spacings
of the Debye-Scherrer peaks in the sample must be loaded into  the program.
This file is then used to carry out the camera length calibration on measured data from the calibration sample.
The 2D diffraction patterns are then integrated into 1D diffraction patterns using these
calibration constants. Conversion to $F(Q)$\cite{egami;b;utbp12} and
then to $G(r)$\cite{egami;b;utbp12} and structural modeling was done using the home
written software packages PDFgetE (unpublished) and Srfit~\cite{juhas;.14} respectively.

\subsection{ePDF data collection protocols}
\label{sec;dcps}
We collected ePDF data using different protocols to test the reliability of different approaches as well as the effect on the resulting quantitative PDF of different measurement aberrations.

\subsubsection{Protocol 1: parallel illumination}

\par Step 1. Load the calibration sample.
\par Step 2. In image mode, adjust objective focus to standard focus or $DV=0$. Focus TEM image by changing sample height to the eucentric height. Insert a selected-area aperture (SAA) to select the area of interest.
\par Step 3. In diffraction mode, insert UOA and focus it by changing diffraction focus.
\par Step 4. In diffraction mode, remove the objective aperture and focus the diffraction spots or rings by adjusting the brightness setting. Record electron diffraction patterns. (Because the UOA is located at the diffraction plane, a parallel illumination condition is assumed, when the UOA is focused in Step 3.)
\par Step 5. Unload the calibration sample and load the real specimen.
\par Step 6. Repeat Step 2 for the real specimen
\par Step 7. Diffraction focus remains untouched, and repeat step 4 for the real specimen

In our case, we collected 200 frames of 0.4~s for electron diffraction to obtain sufficient statistics in the high-$Q$ region of the
diffraction pattern. However, in general the data collection time per frame should be determined by considering the chemical composition, thickness of the
specimen and the threshold intensity of the CCD. The total number of
frames should then be determined depending on the required statistics in the high-$Q$ region of $F(Q)$.  A beam stop was not used for these measurements due to
possible distortions of the diffraction pattern caused by charging of
the beam stop.~\cite{abeyk;zk12}

\subsubsection{Protocol 2: non-parallel illumination}
In this protocol, the LOA is focused in the
diffraction mode instead of the UOA in Step 3, which gives a convergence angle  $\sim0.75$~mrad on our instrument. All other steps in protocol 1 are followed. This is actually the standard procedure for most TEM measurements but may introduce problems for an ePDF measurement which we want to test.

\subsubsection{Protocol 3: introducing a large deviation from the eucentric height}
Step 1 - 5 in Protocol 1 are followed.  In Step 6, the sample height is set to deviate 100~$\mu$m from
the eucentric height, which is compensated by adjusting
image focus by -100~$\mu$m. However, the diffraction focus is not changed and Step 7 is performed.
This is done to test the impact on the calibration of a non-optimal operator procedure.

\subsubsection{Protocol 4: changing diffraction focus}
Step 1-6 in protocol 1 are followed.  However, in Step 7, diffraction focus is adjusted to focus the
LOA, leading to a change in camera length. Our intention is to simulate the extent of the aberration introduced were operators to focus diffraction spots or rings by arbitrarily changing the brightness and diffraction
focus instead of by fixing the diffraction focus and adjusting brightness.

\section{Results and Discussion }

\subsection{Evaluation of the calibration samples}

\subsubsection{Au nanoparticles}
\begin{figure}[tbh]
\center
\includegraphics[scale=.31, angle=0]{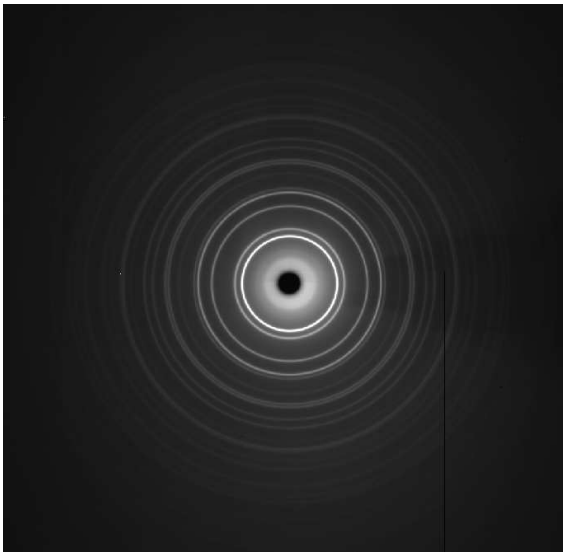}
\caption {X-ray diffraction pattern of the 10~nm diameter Au nanoparticle sample}
\label{fig:Au_XRD}
\end{figure}
\begin{figure}[tbh]
\raggedright
	\begin{minipage}[t]{150pt}
	\includegraphics[scale=0.24]{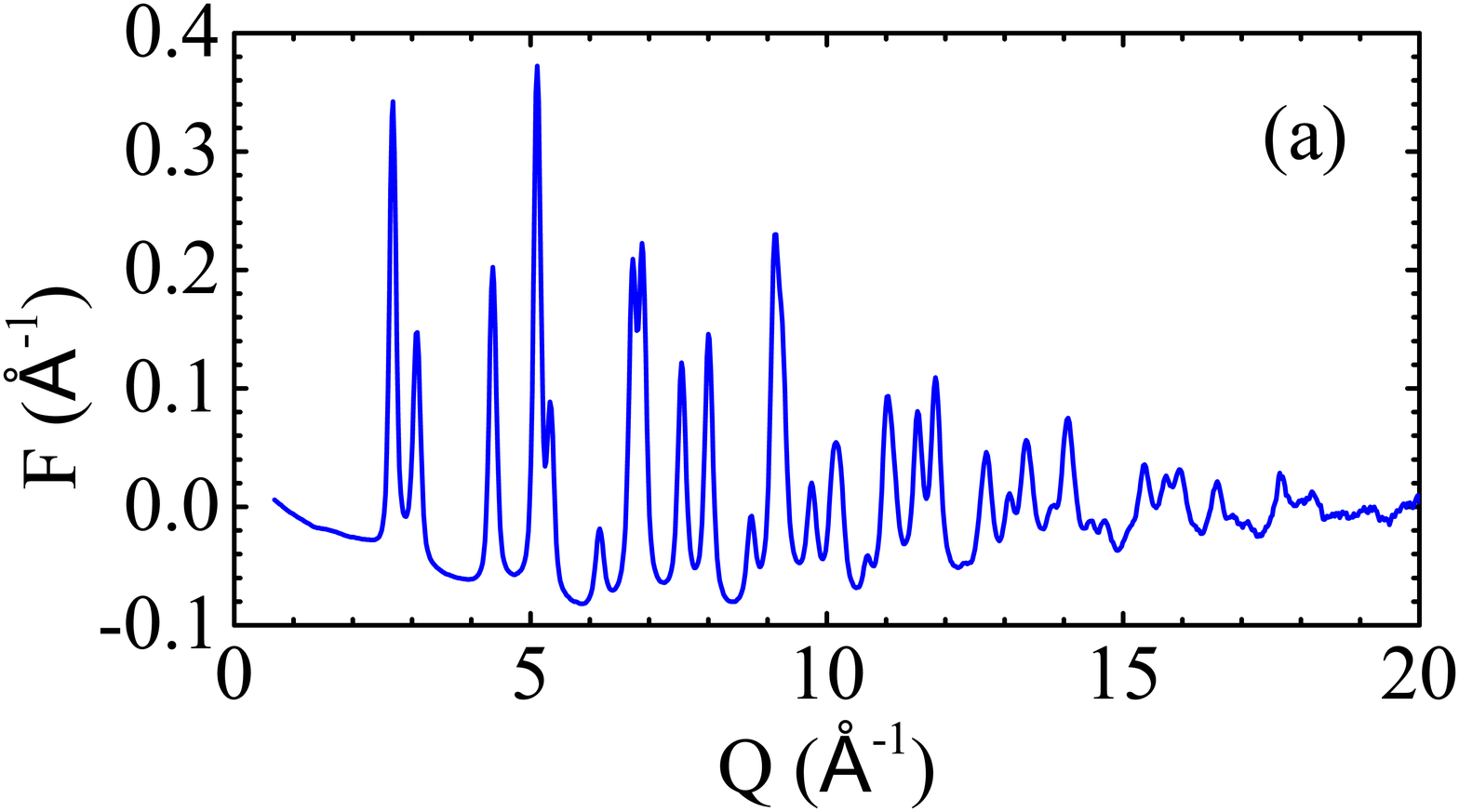}
	\end{minipage}
\hfill
	\begin{minipage}[t]{150pt}
	\includegraphics[scale=0.24]{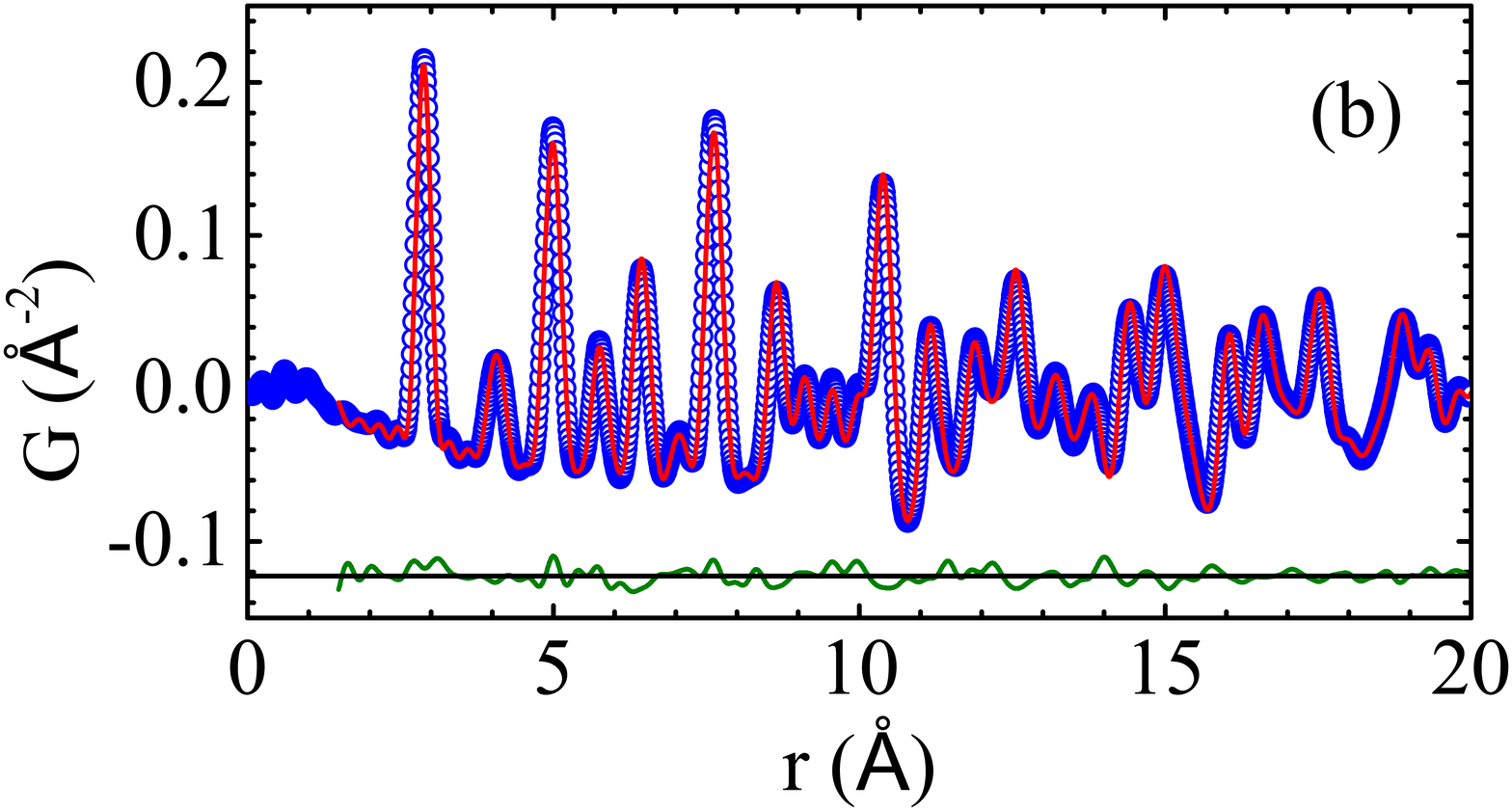}
	\end{minipage}
\caption {(a)Reduced X-ray structure function, $F(Q)$,
  of NIST standard Au nanoparticles calculated from the integrated
  2D XRD pattern in Figure~\ref{fig:Au_XRD}. The resulting PDF, $G(r$), is shown as blue symbols in
  (b).  The best-fit PDF from an fcc structural model is plotted in red with a difference curve offset below.}
\label{fig:Au_xFqGr}
\end{figure}
First we characterized the NIST standard 10~nm Au
nanoparticles (RM8011) using the xPDF
technique to have a baseline reference for the ePDF analysis.
These came suspended in solution. Attempts to perform
X-ray diffraction (XRD) measurements on the as-received solution in 1~mm Kapton capillaries did not produce a
strong enough signal due to the low particle density in the solution. Therefore, the
as-received solution was centrifuged to trap nanoparticles at
the bottom of a vial. The solution at the top of
the vial was carefully removed from the tube leaving (1-2)~ml at the bottom with
the sedimented nanoparticles which was filled into a 1~mm
Kapton capillary for the XRD measurements.  The 2D XRD pattern
from the concentrated solution is shown in
Figure~\ref{fig:Au_XRD}. This pattern indicates a very good powder
average. The reduced structure function
$F(Q)$ calculated from the integrated 1D diffraction pattern of Au is shown in
Figure~\ref{fig:Au_xFqGr}(a), which also indicates the good statistics on the data up to
$Q_{max}=20$~{\AA$^{-1}$}. The resulting xPDF is shown in Figure~\ref{fig:Au_xFqGr}(b)
with a best-fit PDF from the FCC Au structural model overlaid.
The xPDF experimental
resolution function, $Q_{damp}$, 
was obtained in the usual way by modeling
a bulk Ni standard xPDF and this is fixed in subsequent modeling.

We found very good agreement between the experimental and calculated PDFs from the Au nanoparticles.
The lattice constant, particle size and
atomic displacement parameter (ADP) of the Au nanoparticles were obtained from the xPDF
analysis, which will be used as a reference for ePDF analysis using the same standard sample.
A calibration file of $d$-spacings
of the Au nanoparticles was prepared to be used in Fit2D to
obtain the exact camera length used in the ePDF measurement. The calculated $d$-spacings
of Au that were used in the Fit2D calibration file are given in Table~\ref{tab:Au_d}.
%
\begin{table}[tbh]
\caption{The Miller Indices (h,k,l) and the corresponding lattice $d$-spacings
  of Au obtained from the xPDF analysis of the sample.}%
\label{tab:Au_d}
\begin{ruledtabular}
\begin{tabular}{lll}
Miller Indices hkl         &Lattice spacing $d$ (\AA)   \\
\hline
111                             & 2.3534               \\
200                             & 2.0381                \\
220                             & 1.4411              \\
311                             & 1.2290              \\
222                             & 1.1767              \\

\end{tabular}
\end{ruledtabular}
\end{table}

\begin{figure}[tbh]
\center
\includegraphics[scale=.23, angle=0]{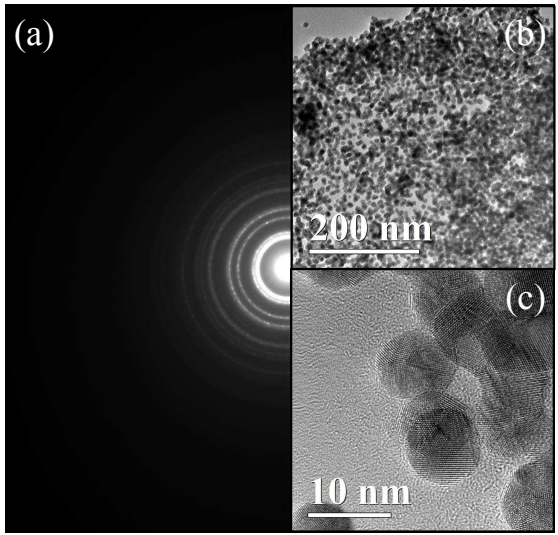}
\caption {(a)A typical electron diffraction pattern of the 10~nm diameter Au nanoparticle sample. Panel (b) and (c) shows the TEM image at low and high magnification, respectively.}
\label{fig:Au_2D}
\end{figure}

Attempts to make an ePDF calibration sample by placing a drop of the
as-received solution on a TEM grid failed, again due to the low nanoparticle
density. However, the
TEM specimen prepared by centrifuging the as-received solution
in a vial with a TEM grid placed at the bottom gave a very good signal
producing the 2D electron diffraction
pattern shown in Figure~\ref{fig:Au_2D}(a). This electron diffraction pattern shows a good powder average indicating
a uniform 3D distribution of nanoparticles on the TEM grid. The TEM
images shown in  Figure~\ref{fig:Au_2D}(b) and (c) indicate a uniform distribution of particle size and shape.
\begin{figure}[tbh]
\raggedright
	\begin{minipage}[t]{150pt}
	\includegraphics[scale=0.24]{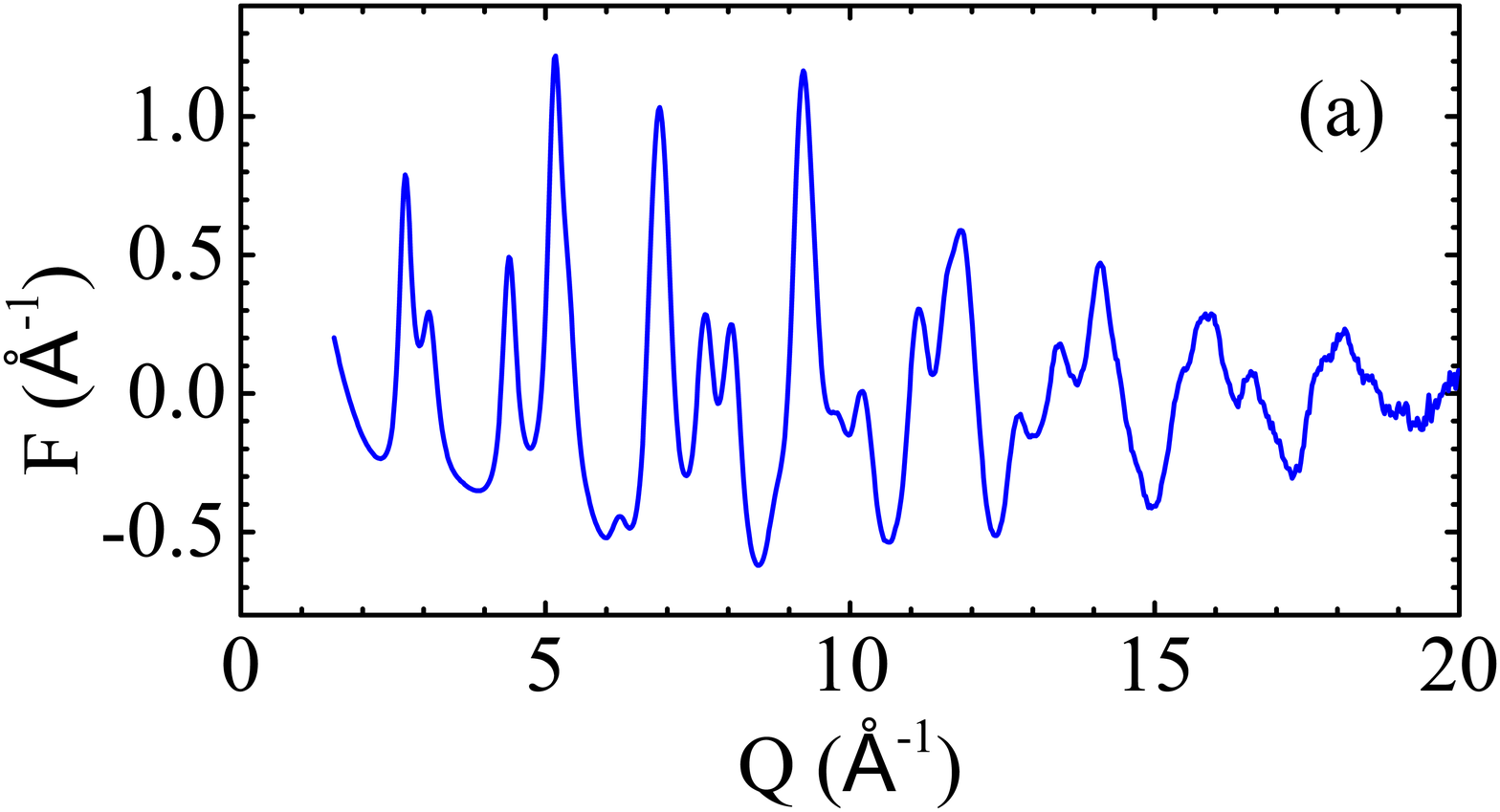}
	\end{minipage}
\hfill
	\begin{minipage}[t]{150pt}
	\includegraphics[scale=0.24]{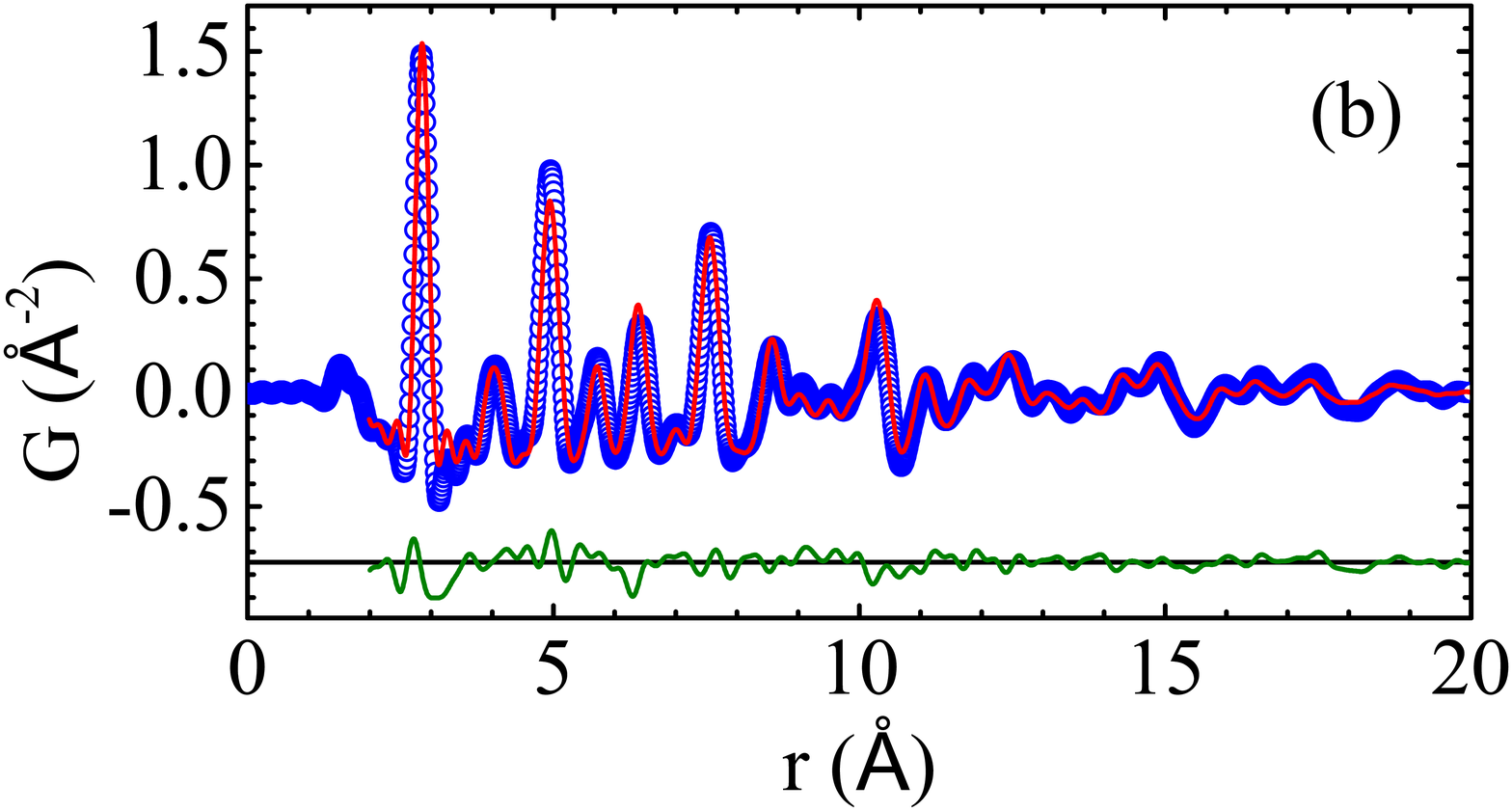}
	\end{minipage}
\caption {(a) Reduced electron structure function, $F(Q)$,
  of NIST standard Au nanoparticles calculated from the integrated 2D ED pattern
  in Figure~\ref{fig:Au_2D}. The resulting PDF, $G(r)$,
  is shown as blue symbols in (b). The best-fit PDF from an fcc structural model is plotted in red with a difference curve offset below.}
\label{fig:Au_FqGr}
\end{figure}

The reduced electron structure function $F(Q)$, calculated from the integrated 1D electron diffraction pattern of Au given in
Figure~\ref{fig:Au_FqGr}(a), shows data with very good statistics up to
$Q_{max}=20$~{\AA$^{-1}$}. The resulting PDF is shown using blue symbols in Figure~\ref{fig:Au_FqGr}(b). Plotted on top
is the ePDF calculated from the structural model for FCC Au.  The agreement is very good.
The ePDF refinement results are compared with the xPDF results in
Table~\ref{tab:Table_Au_xe}.
\begin{table}[tbh]
\caption{Comparison of refined parameters of Au obtained from  xPDF and ePDF modeling. The structure model is the fcc bulk Au structure, space group $Fm\bar{3}m$. $Q_{damp}$, obtained from the Ni calibration sample, was fixed during the modeling for xPDF. The particle size, refined from the xPDF modeling, was kept constant during the ePDF modeling. The description of fit parameters is listed in the footnote.}
\label{tab:Table_Au_xe}
\begin{ruledtabular}
\begin{tabular}{llllll}
  Au                                                                   & xPDF                           & ePDF \\
\hline
$a$ (\AA)                                                              & 4.076                          & 4.076 \\
$U_{Au}$ (\AA$^{2}$)\footnote{atomic displacement parameter(ADP)}     & 0.099                          & 0.011  \\
$Q_{damp}$ (\AA$^{-1}$)\footnote{instrument resolution parameter}     & 0.043$^f$                      & 0.095  \\
$\delta_2$ (\AA$^{2}$)\footnote{peak sharpening parameter}            & 3.87                           &  5.71  \\
$Q_{broad}$ (\AA$^{-1}$)\footnote{peak broadening parameter}          & 0.057                          & 0.039\\
spdiameter (\AA)\footnote{particle size}                              & 100.6                          & 100.6\footnote{fixed to the xPDF value during the refinement}\\
$R_{w} $(\%)                                                           & 9                              &  18   \\
\end{tabular}
\end{ruledtabular}
\end{table}

The good agreement between the lattice parameters obtained from
the xPDF and ePDF analysis is expected since we used the $d$-spacings 
obtained from the xPDF measurement to calibrate the camera length. However, there is also good agreement for
the ADPs indicating minimal multiple
scattering effects in this sample~\cite{abeyk;zk12}.

Using $Q_{damp}$ obtained from the Ni calibration sample, we obtained a particle size of 10.06~nm from the xPDF measurement, in good
agreement with the nominal size of 10~nm reported by NIST. 

In general, the PDF signal damps at high-$r$ due to two factors,
the experimental $Q$-space 
resolution and the average particle size in a nanoparticle
sample~\cite{egami;b;utbp12}. In X-ray and neutron PDF
studies, the bulk standard sample measurement used to obtain the
sample-to-detector distance can also be used to determine the
instrument resolution function in the PDF analysis, $Q_{damp}$, by
assuming an infinite particle size for the crystalline calibrant. However, it is not
possible to use a bulk sample as the calibration standard in ePDF
studies due to the enhanced multiple scattering signal. In this situation, an electron diffraction pattern from a standard nanomaterial (with a known particle size)
is a better choice to calibrate the camera length and the PDF instrument
resolution function, $Q_{damp}$. Here we fully characterized the NIST standard 10~nm Au
nanoparticles using xPDF, and these values were used
to calibrate the camera length and $Q_{damp}$ in ePDF experiments where the same sample was used as a calibrant.
These values are presented in Table~II in the xPDF column.  The procedure to follow is to measure the NIST-standard Au
nanoparticle sample in the TEM and do the camera length calibration. Then, during ePDF modeling, fix spdiameter to 100.6~\AA, the number refined from the xPDF analysis, but allow $Q_{damp}$ to vary.  The refined value of $Q_{damp}$
should then be fixed and used in subsequent refinements on ePDFs of the real samples.  The parameter spdiameter can then be allowed to vary and should give an accurate determination of the particle size (or, more accurately, size of the coherent structural domain~\cite{egami;b;utbp12}) in the sample.


\subsubsection {Evaporated Aluminum}

Another convenient specimen for TEM camera-length calibration is an evaporated aluminum film deposited on G400, 400 square mesh copper
Gilder grid,  from Ted Pella,~Inc. 
A list of principal lattice spacings necessary to make a software
calibration file is provided with the sample by the
manufacturer and reproduced for completeness in Table~\ref{tab:Ev_Al_dspacings}).
\begin{table}[tbh]
\caption{Miller Indices ($h,k,l$) and corresponding lattice $d$-spacings
  of evaporated aluminum film. The average lattice parameter is 4.0494~\AA}
\label{tab:Ev_Al_dspacings}
\begin{ruledtabular}
\begin{tabular}{lll}
Miller Indices hkl         &Lattice spacing $d$ (\AA)   \\
\hline
111                              & 2.3380              \\
200                              & 2.0240              \\
220                              & 1.4310              \\
311                              & 1.2210              \\
222                              & 1.1690              \\
\end{tabular}
\end{ruledtabular}
\end{table}

\begin{figure}[tbh]
\center
\includegraphics[scale=.255, angle=0]{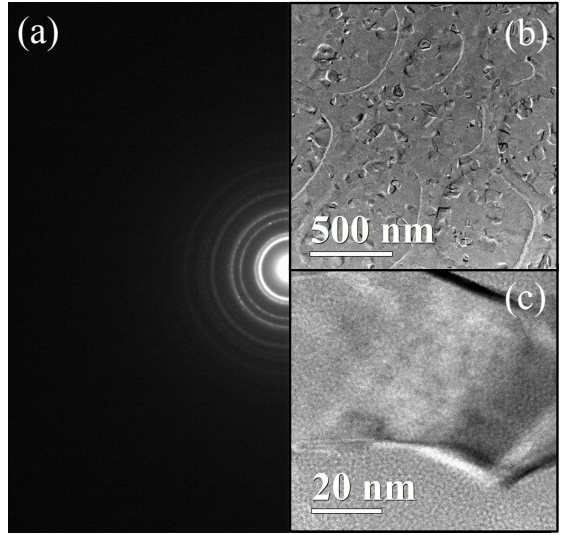}
\caption { (a) An averaged electron diffraction pattern from 30 different areas of the evaporated aluminum film to have better sampling in orientation. Panels (b) and (c)
  shows the TEM image of the film at low and high magnification, respectively}
\label{fig:EvAl_2D}
\end{figure}

To investigate whether the commercial Al films are accurate calibrants for camera length calibration, three Al-film samples were tested. The TEM camera length was first calibrated using the NIST-standard Au sample following Protocol 1, which was then removed. Three Al film samples, from the same order received from the manufacturer, were subsequently measured one by one using the same protocol. It was noted that the electron diffraction patterns of the Al films were very spotty when the similar beam size ($\sim2$~$\mu$m) was used for the NIST Au nanoparticle sample. Therefore, 30 electron diffraction patterns recorded from different areas were averaged to obtain a better powder average. The averaged electron diffraction pattern is shown in Figure~\ref{fig:EvAl_2D}(a), and the TEM image at low and high magnification is shown in (b) and (c), respectively.
The camera length was refined to be 79.61~mm 
 for the Au nanoparticle sample, while the three evaporated Al samples gave 79.15~mm, 78.74~mm and 79.48~mm, respectively. We therefore found a significant variability in the camera length obtained from different Al-films. The evaporated Al samples are inexpensive and convenient to use, but appear to be giving camera lengths with an accuracy no better than the 1~\%\ level.  If higher accuracy than this is needed, then the particular Al film used should itself be calibrated, or NIST-standard Au should be used.  A further drawback of using the Al films as an ePDF calibration sample for more accurate PDF modeling is that the particle size is not known and therefore we cannot get a reliable value for $Q_{damp}$.  For the most accurate work we recommend use of the NIST standard nanoparticle Au sample as a calibrant.

\subsection{Evaluation of data collection protocols}
\begin{figure}[h!]
\center
\includegraphics[scale=.31, angle=0]{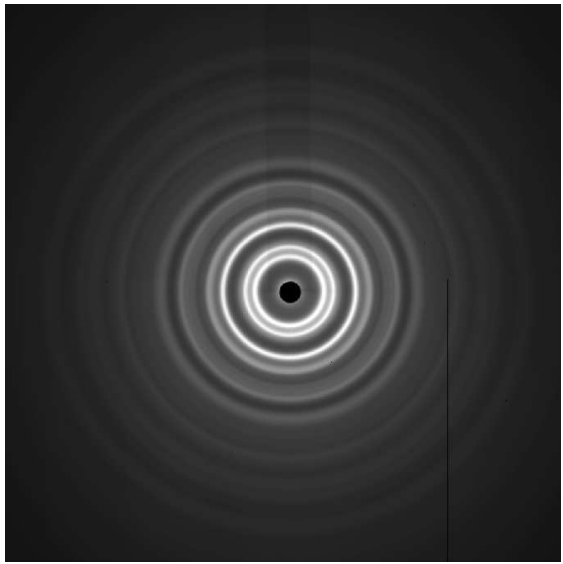}
\caption {Representative 2D image of the X-ray diffraction pattern from the SnO$_{2}$ nanoparticle specimen.}
\label{fig:SnO2_2DXRD}
\end{figure}
\begin{figure}[t]
\raggedright
	\begin{minipage}[t]{150pt}
	\includegraphics[scale=0.24]{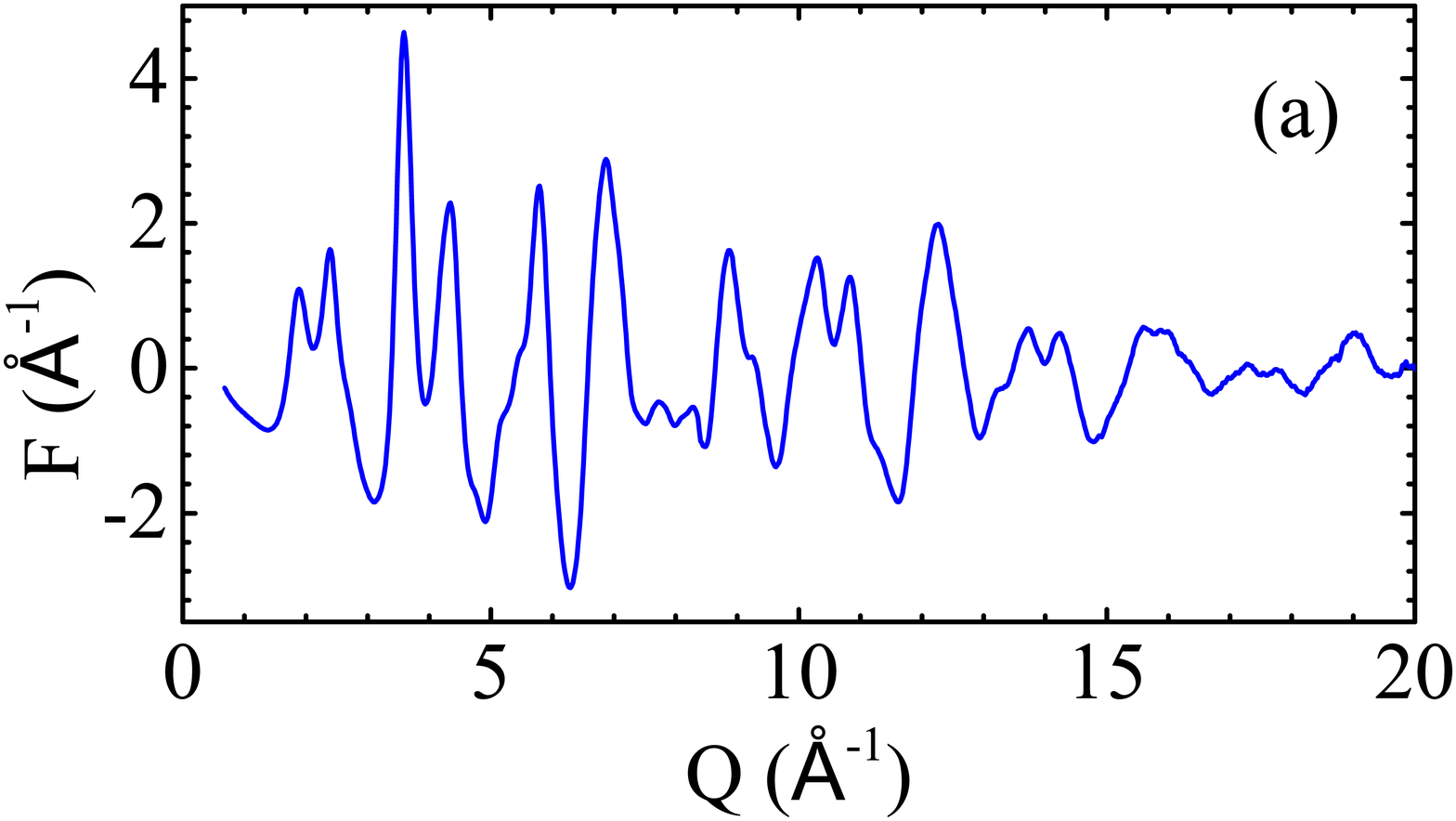}
	\end{minipage}
\hfill
	\begin{minipage}[t]{150pt}
	\includegraphics[scale=0.24]{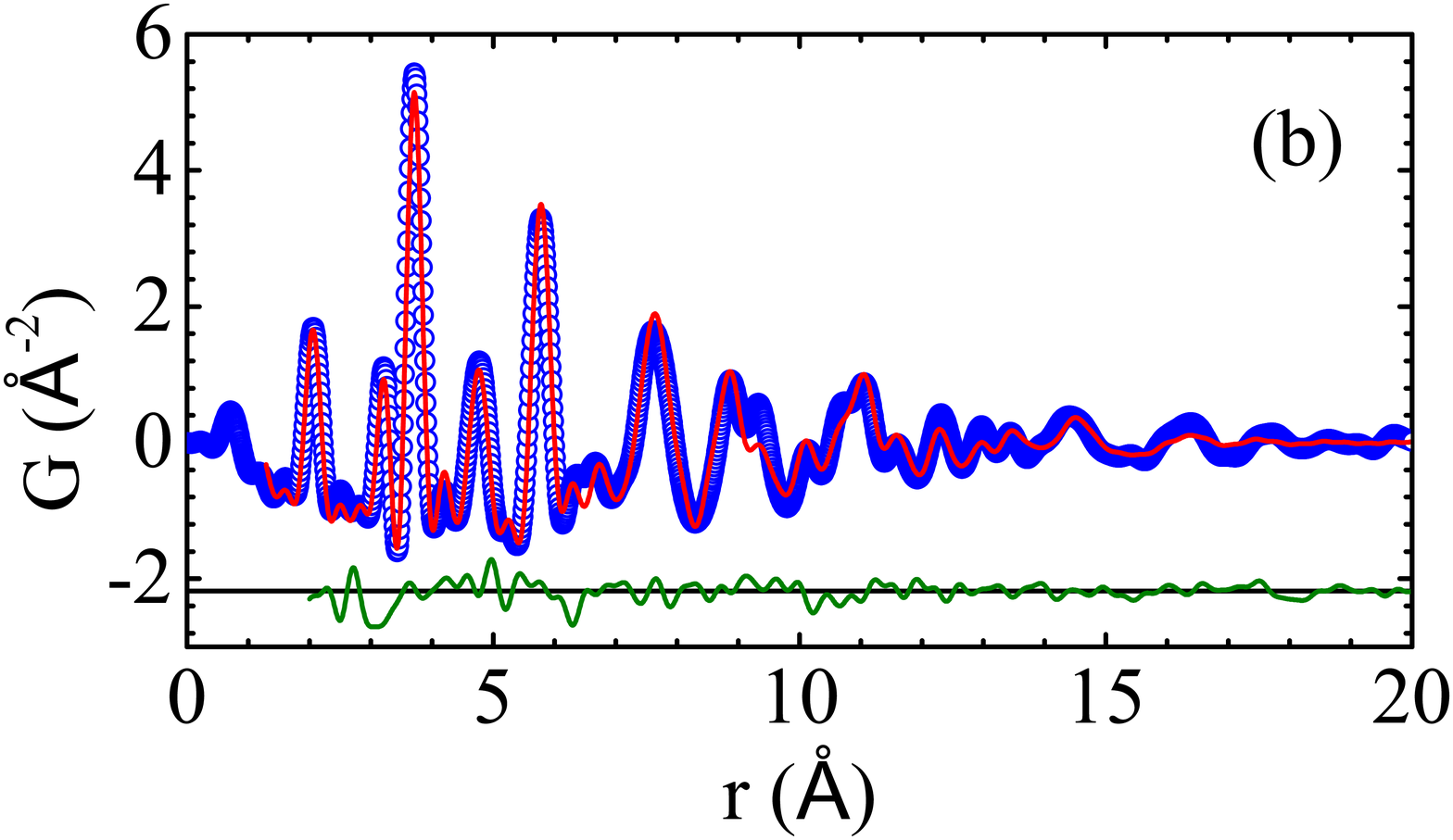}
	\end{minipage}
\caption {(a) Reduced X-ray structure function, $F(Q)$, from the SnO$_2$ NP sample calculated from the integrated 2D XRD pattern in Figure~\ref{fig:SnO2_2DXRD}(a).  The resulting PDF, $G(r)$, is shown as blue symbols in (b). The best-fit PDF from a structural model is plotted in red with a difference curve offset below. }
\label{fig:SnO2_xFqGr}
\end{figure}

Here we compare the results obtained using different data collection protocols described in
Sec.~\ref{sec;dcps}.  A commercial 
SnO$_{2}$ nanoparticle sample (Nano-Oxides, Inc, catalog number: 18-025) was used for this purpose. This
sample was first characterized by using the xPDF technique to have a reference
in the ePDF study. 
The 2D X-ray diffraction pattern from the SnO$_{2}$ nanoparticle sample
is shown in Figure~\ref{fig:SnO2_2DXRD}.  The rings are smooth indicating a good powder average, and broad suggesting a nanocrystalline sample.
The resulting $F(Q)$ and $G(r)$ are shown in
Figure~\ref{fig:SnO2_xFqGr}, (a) and (b) respectively.  The xPDF from the best-fit model is plotted in red over the data with a difference curve offset below.  The agreement is acceptable for a nanocrystalline sample.
\begin{figure}[h!]
\center
\includegraphics[scale=.25, angle=0]{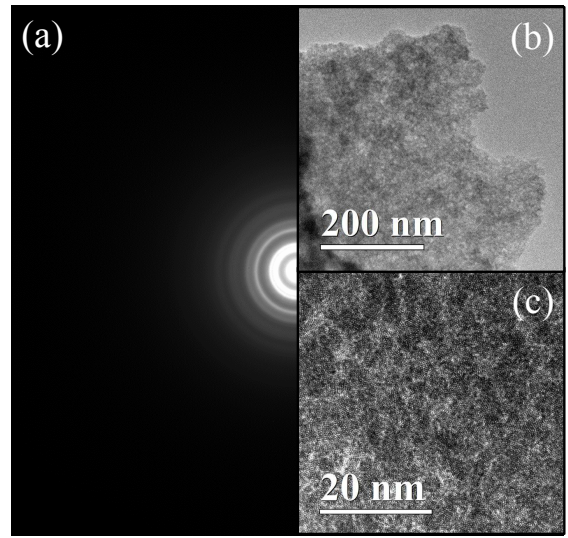}
\caption {(a) A typical electron diffraction
  pattern of the SnO$_{2}$ nanoparticle specimen. Panels (b) and (c)
  shows the TEM image of the SnO$_{2}$ sample at low and high magnification, respectively}
\label{fig:SnO2_2D}
\end{figure}
\begin{figure}[t]
\raggedright
	\begin{minipage}[t]{150pt}
	\includegraphics[scale=0.24]{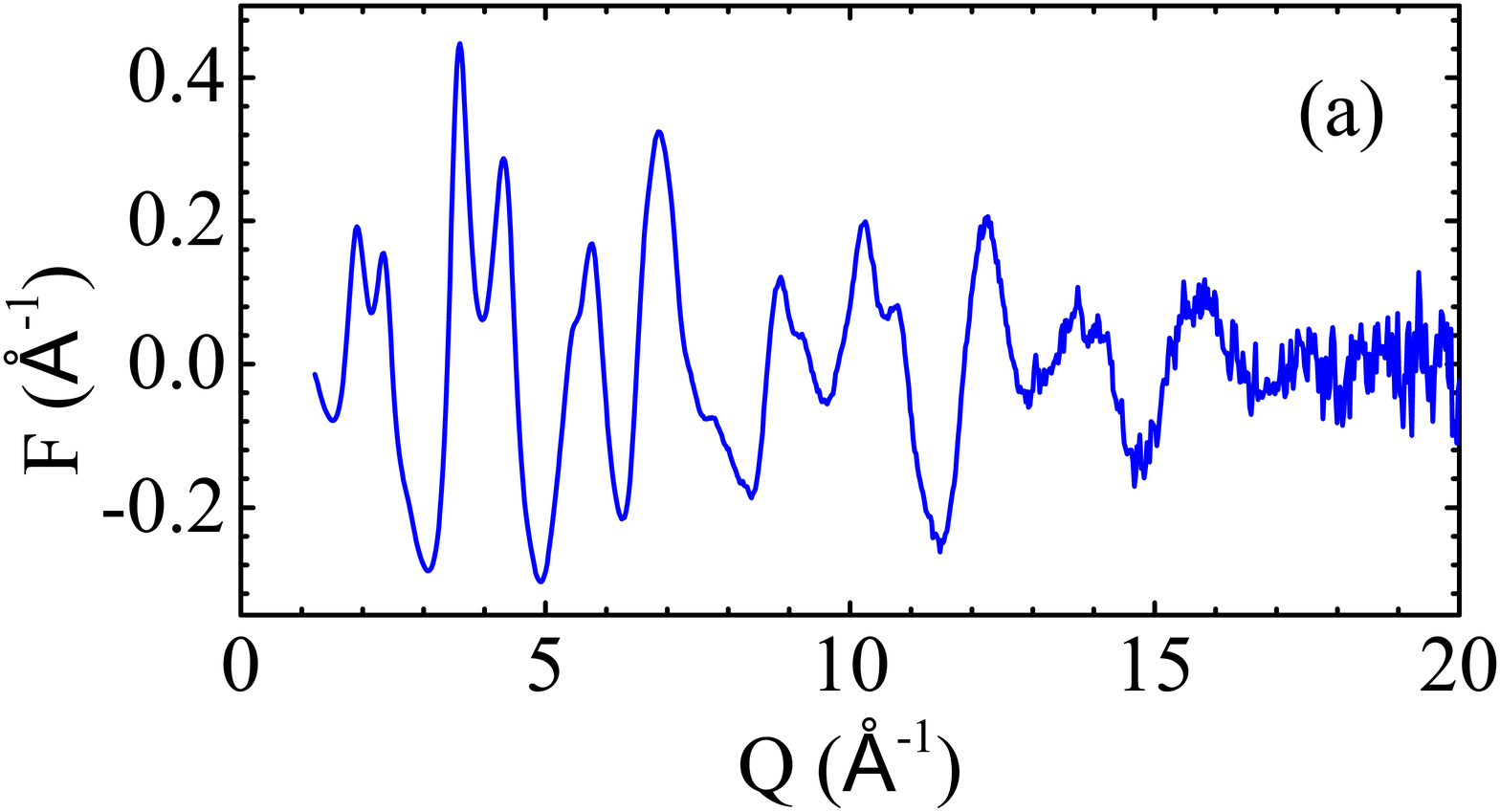}
	\end{minipage}
\hfill
	\begin{minipage}[t]{150pt}
	\includegraphics[scale=0.24]{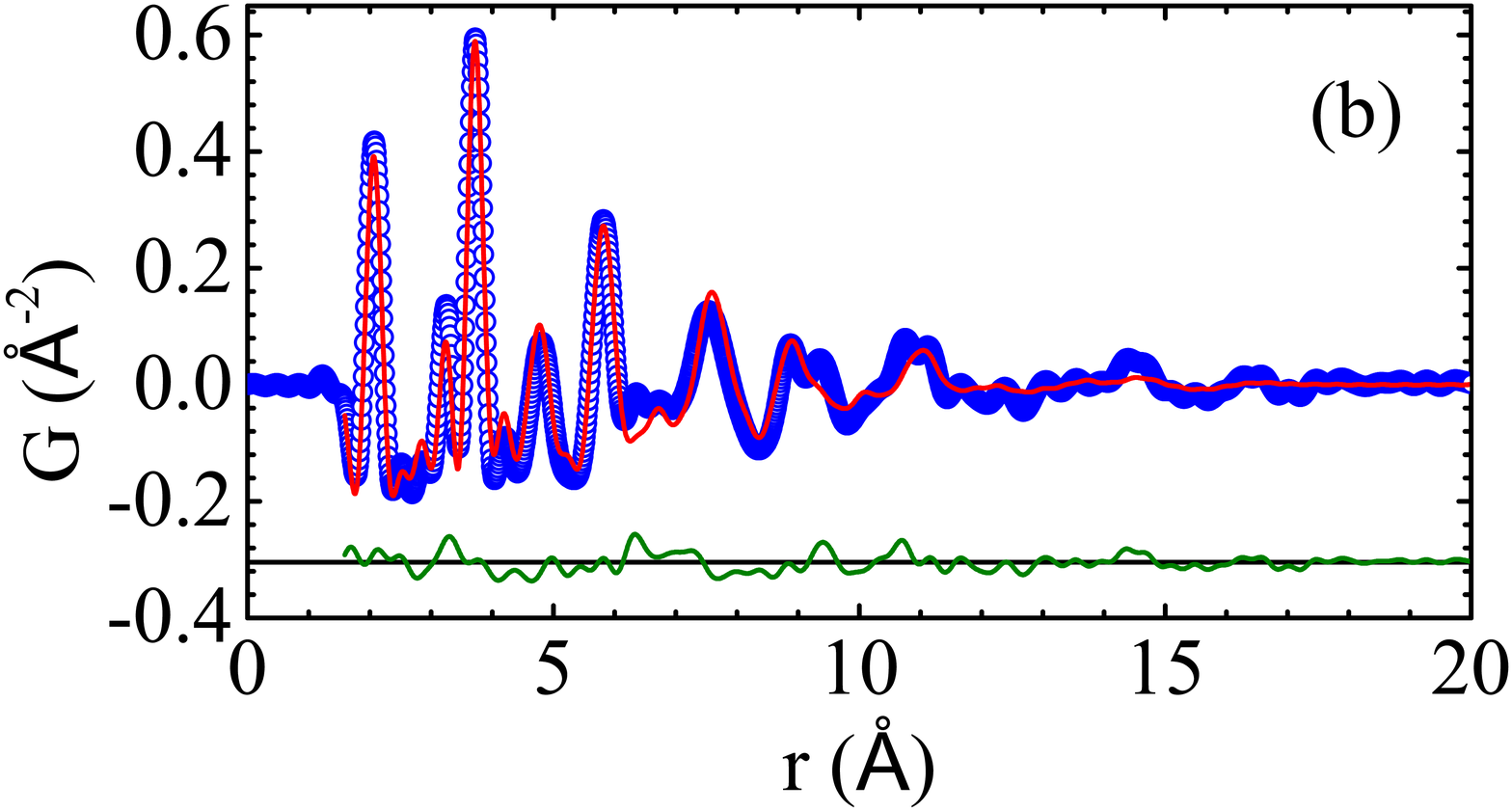}
	\end{minipage}
\caption {(a)Reduced electron structure function, $F(Q)$, from the SnO$_2$ nanoparticle sample calculated from the integrated 2D electron diffraction pattern in Figure~\ref{fig:SnO2_2D}(a). The resulting PDF, $G(r)$, is shown as blue symbols in (b). The best-fit PDF from a structural model is plotted in red with a difference curve offset below. }
\label{fig:SnO$_2$_FqGr}
\end{figure}


Material from the same sample was then studied using electron diffraction.
The 2D electron diffraction pattern and real-space TEM images from the SnO$_{2}$ nanoparticle sample are shown in Figure~\ref{fig:SnO2_2D}.  The 2D electron diffraction pattern is smooth, indicating a good powder average while
the images indicate a uniform particle size and shape distribution. The resulting $F(Q)$ and ePDF is shown in Figure~\ref{fig:SnO$_2$_FqGr}(a) and (b), respectively, again with the best-fit PDF from the structural modeling plotted over the top of ePDF.
The similarity to the xPDF in Fig.~\ref{fig:SnO2_xFqGr} and the good agreement of the structural models indicate the good quality of the ePDF data.

The values of refined parameters from fits to the xPDF as well as ePDFs obtained with different protocols is reproduced in Table~\ref{tab:protocals}. For both xPDF and ePDF modeling, the instrument resolution parameter, $Q_{damp}$, was fixed to the value refined from the calibration sample, respectively. For a more clear comparison of lattice parameters between the ePDF and xPDF results, the differences with respect to the xPDF-refined values are shown as a percentage for different protocols of ePDF, respectively. Protocol 1-3 gave deviations within 0.4\% and 1.3\% from the xPDF values for lattice parameter $a$ and $c$, respectively,  while differences of 1.0\% and 2.6\% were found for $a$ and $c$, respectively, in Protocol 4. Overall, protocol 4 resulted in bigger differences from the xPDF reference than other protocols.

The only difference between Protocol 1 and 2 is the illumination condition with parallel illumination in Protocol 1 and non-parallel illumination in Protocol 2. Therefore, once the camera length is calibrated in the same condition, non-parallel beam illumination should still lead to correct results. Moreover, as tested in Protocol 2, with all other conditions fixed, a slightly convergent beam could improve the signal statistics at high $Q$ to some extent, since a convergent beam leads to a larger electron beam current when an SAA is used. Protocol 3 also gave similar results with Protocol 1 and 2, which indicates that deviation in sample height at least up to 100~$\mu$m has little or no impact on quantification of electron diffraction in the our test circumstance. However, this condition led to large distortion of real-space images, affecting correct estimation on size and shape of nanoparticles, which is not recommended. Protocol 4 simulates the wrong operation that diffraction focus is adjusted to focus diffraction spots or rings after the real specimen is loaded. Our settings gave a change of 1.6~\% in camera length, which resulted in a larger deviation in lattice parameters from the xPDF reference in Protocol 4. This deviation is directly correlated with the change in camera length. Therefore, change in diffraction focus during TEM operation should be strictly avoided.

Multiple electron scattering can give rise to complications of quantifying electron diffraction patterns, especially when the sample is thick and oriented on its zone axis. In ePDF experiments, this multiple scattering effect is expected to be reduced due to the small nanoparticle size and randomly distributed orientations. However, to fully understand the impact of multiple scattering on randomly oriented nanoparticle samples, further experimental studies together with theoretical simulation will be conducted in the future.

\begin{table*}[htp!]
\caption{Refined parameters from the modeling of the SnO$_{2}$ PDF data. The columns contain results from xPDF data
and from ePDF data collected with the different protocols described in Sec.~\ref{sec;dcps}.  The structural model is
the bulk SnO$_{2}$, space group $P4_{2}/mnm$. Sn is on position (0,0,0). Oxygen positions ($x(O)$,$y(O)$,0), with $x(O)=y(O)$ from symetry,
 were refined during the modeling. $Q_{damp}$ was fixed during the modeling for xPDF and ePDF. In the ePDF columns, the percentages in the brackets next to lattice parameters are the differences with respect to the values obtained from xPDF modeling. } 
\label{tab:protocals}
\begin{ruledtabular}
\begin{tabular}{llllll}
SnO$_{2}$                 & xPDF                                          & protocol 1     & protocol 2     &protocol 3      & protocol 4     \\
\hline
$a$ (\AA)                 & 4.751                                         &  4.75(0~\%)    & 4.73(-0.4~\%)  & 4.76(0.2~\%)   & 4.80(1.0~\%)  \\
$c$ (\AA)                 & 3.187                                         &  3.23(1.3~\%)  & 3.21(0.7~\%)   & 3.23(1.3~\%)   & 3.27(2.6~\%)  \\
$x(O)$                    & 0.3004                                        &  0.2970        & 0.2953         & 0.2956         & 0.2960         \\
$U_{Sn}$ (\AA$^{2}$)      & 0.003                                         &  0.001         & 0.001          & 0.003          & 0.001          \\
$U_{O}$ (\AA$^{2}$)       & 0.03                                          &  0.01          & 0.01           & 0.02           & 0.01           \\
$Q_{damp}$ (\AA$^{-1}$)   & 0.043\footnote{fixed during the refinement}  &  0.095$^a$     & 0.095$^a$      & 0.095$^a$      & 0.095$^a$            \\
$\delta_2$ (\AA$^{2}$)    & 3.4                                           &  5             & 5              & 5              & 6              \\
$Q_{broad}$ (\AA$^{-1}$)  & 0.14                                          &  0.4           & 0.5            & 0.2            & 0.5            \\
spdiameter (\AA)          & 21                                            &  22            & 24             & 25             & 25        \\
$R_{w}$ (\%)              & 18                                            &  24            & 25             & 25             & 26          \\
\end{tabular}
\end{ruledtabular}
\end{table*}

\section{Conclusions}

To quantify ePDF measurements, the NIST-standard Au nanoparticle sample is recommended over the commercial Al-film sample, because the Au sample provides more reliable calibration of camera length and the uniform size distribution of the Au nanoparticles can be used to determine the instrument resolution, $Q_{damp}$, in TEM. Four ePDF protocols, including common/possible operational mistakes, were tested by comparing the refined parameters of the SnO$_2$ nanoparticle sample in ePDF with the xPDF results. Protocol 1 and 2 with parallel and slightly convergent illumination, respectively, showed accuracy within 0.4~\% and 1.3~\% for determining lattice parameter $a$ and $c$, respectively. Protocol 3 with incorrect sample height also led to very similar results as in protocol 1 and 2. However, real-space images were significantly distorted, which made estimation on nanoparticle size and shape difficult. Hence, Protocol 3 is not recommended. Diffraction focus was adjusted in Protocol 4 after the Au calibration diffraction was recorded, which directly led to a change in camera length and should be strictly avoided. Overall, Protocol 1 and 2 are recommended.

\acknowledgments{ This work
was carried out as part of the Flucteam project at BNL supported by the US Department of Energy, Office of Science, Office of Basic Energy Sciences (DOE-BES) through account DE-AC02-98CH10886. Use of the National Synchrotron Light Source, Brookhaven National Laboratory, was supported by the DOE-BES under Contract No. DE-AC02-98CH10886.
}


\end{document}